\documentstyle[12pt,aaspp4]{article}
\def \aox          {{$\alpha_{ox}$}}
\def \cmsq           {\hbox{cm$^{-2}$}}
\def \deg          {\ifmmode ^{\circ}\else $^\circ$\fi}  
\def \etal         {{\it et~al.} }

\def \kms          {\rm{\hbox{km s$^{-1}$}}}
\def \lam          {$\lambda$}
\def \Lya          {\hbox{Ly$\alpha$}}

\def \mum          {\hbox{$\mu$m}}
\def \pcc           {\hbox{cm$^{-3}$}}
\def \zaz          {{$z_a\kern -1.5pt \approx\kern -1.5pt z_e$}}
\def \zllz         {{$z_a\kern -3pt \ll\kern -3pt z_e$}}
\def \Msun         {\rm{\hbox{M$_{\odot}$}}}           
\def \Zsun         {\rm{\hbox{Z$_{\odot}$}}}           

\begin{document}
\renewcommand{\baselinestretch}{2}
\title
{\huge
Metal Abundances and Ionization \\
in QSO Intrinsic Absorbers}

\renewcommand{\baselinestretch}{1}
\smallskip
\author
{\large
Fred Hamann}

\affil
{Center for Astrophysics and Space Sciences \\
University of California, San Diego; La Jolla, CA 92093-0111 \\
Internet: fhamann@ucsd.edu}
%
%
%
%
\bigskip

\begin{abstract} 
\small

This paper provides a general reference for deriving the 
ionizations and metal abundances for absorption-line clouds 
in photoionization equilibrium with a QSO spectrum. The 
results include a quantitative assessment of the theoretical 
uncertainties and firm lower limits on the metallicities 
when no ionization constraints are available. 
The calculations applied to the best available column densities 
for QSO intrinsic absorbers support previous studies; 
there is strong evidence for super-solar 
metallicities in broad absorption line (BAL) and at least some 
``associated'' (\zaz ) absorption regions. 
Even in cases where the level of ionization is unknown (for example,  
when only H I and C IV lines are measured), firm lower limits on the 
metal abundances (C/H) imply $Z$~$\ga$~\Zsun\ for 
typical BALs and some \zaz\ absorbers. These 
results support the independent evidence for high 
metallicities in QSO broad emission-line regions. I argue that  
high gas-phase abundances, up to $\sim$10~\Zsun , can be expected 
in QSOs due to the normal enrichment by stars in the cores of 
young, massive ($\ga$10$^{11}$~\Msun ) galaxies.

The measured column densities in intrinsic absorbers also indicate that the 
gas is usually optically thin in the H I Lyman continuum out to energies 
above at least the C IV and N V ionization edges (i.e. $\ga$100 eV). The 
column density ratios often require a range of ionization states 
in the same absorber. Neighboring \zaz\ systems (having 
similar redshifts and often blended spectroscopically) also 
usually require different ionizations or metal abundances. 
If the abundances are similar in neighboring systems, their 
ionization parameters must sometimes differ by more 
than an order of magnitude, requiring a significant range of 
densities and/or distances from the ionizing QSO. These result, and 
the overall levels of ionization, are not consistent with 
single-zone models of the UV line and X-ray continuum (``warm'') 
absorbers. 

\end{abstract}

\keywords{Galaxies: abundances --- Galaxy evolution ---
Quasars: absorption lines --- Quasars: general}

\section{INTRODUCTION}

Metal absorption lines in QSO spectra are important probes 
of galactic evolution and star formation at redshifts up to $z$ $\sim$ 5. 
Most studies have emphasized the cosmologically intervening systems, with 
redshifts much less than the QSO emission redshift (\zllz ), as probes of 
extended structures such as galactic halos and (perhaps) the disks of 
young spiral galaxies. However, 
if QSOs reside in galactic nuclei, there is also 
the possibility of probing galactic nuclear 
environments by measuring absorption (and emission) lines that form 
physically close to the QSOs. Such ``intrinsic'' features could 
yield information about galactic nuclear evolution that is 
not available by other means.  

Intrinsic absorbers include the broad absorption lines 
(BALs) and any of the so-called associated (\zaz ) metal line 
systems that are physically related to QSOs or active galactic 
nuclei (AGNs). 
BALs are understood as forming in high velocity outflows that reach up 
to $\sim$0.1$c$ within $\sim$1 kpc of 
radio-quiet QSOs (Turnshek 1988; Weymann \etal 
1985 and 1991; Stocke \etal 1992; Hamann, Korista \& Morris 1993; 
Barlow \etal 1992; Barlow 1993). 
The \zaz\ systems have much narrower lines and 
could have a variety of origins 
(Weymann \etal 1979). The standard definition of \zaz\ lines in QSOs 
has been that they appear within several thousand \kms\ of the  
emission redshift and are less than a few hundred 
\kms\ wide (cf. Weymann \etal 1979; Foltz \etal 1988; Anderson \etal 1987). 
Some \zaz\ systems meeting these criteria might reside in external 
galaxies, like intervening \zllz\ absorbers, while others are 
physically close to the QSOs. Direct evidence for a 
close physical relationship has come from time-variable 
line strengths. For example, in two QSOs, UM~675 ($z_e$~=~2.15;  
Hamann \etal 1995a) and Q2343+125 ($z_e$~=~2.515; Hamann, Barlow \& 
Junkkarinen 1996a), the variability timescales imply minimum 
space densities above several times 10$^3$ \pcc\ 
and maximum distances of $\sim$1 kpc from the ionizing 
continuum source. More dramatic time-variability, requiring larger 
densities and smaller distances, appears to be common 
in the absorption lines of Seyfert 1 galaxies 
(Ulrich 1988; Voit, Shull \& Begelman 1987; Shull \& Sachs 1993; 
Kriss \etal 1995; Koratkar \etal 1996; Maran \etal 1996). 
Further evidence for the intrinsic nature of \zaz\ lines in QSOs 
has come, ofr example, from observations of (1) partial coverage of 
the background emission 
source(s) (well resolved, optically thick lines with too-shallow 
absorption troughs; Wampler, Bergeron \& Petitjean 1993; 
Petitjean, Rauch \& Carswell 
1994; Hamann, Barlow \& Junkkarinen 1996; Hamann \etal 1997a and 1997b; 
Barlow \etal 1996), and (2) 
smooth and relatively broad line profiles (up to a few hundred \kms\ FWHM) 
that are markedly different from the narrow, often multi-component 
intervening systems (Blades 1988; Hamann \etal 1997a and 1997b; 
Barlow \etal 1996).

Some recent studies also indicate that intrinsic absorbers are generically 
metal rich, with metallicities often above solar (Wampler 
\etal 1993; M\o ller, Jakobsen \& Perryman 1994; Petitjean \etal 
1994; Savaglio \etal 1994; Hamann \etal 1995a and 1997b; 
Tripp \etal 1996; Turnshek \etal 1996; Korista \etal 1996). 
These high metallicities are significantly larger than the 
intervening systems (where typically $Z$~$\la$~0.1~\Zsun ; 
Bergeron 1988; Pettini \etal 1994), but they are consistent with recent
estimates of $Z$ $\ga$~\Zsun\ in the broad emission line regions of 
QSOs (Hamann \& Ferland 1992 and 1993; Ferland \etal 1996; Hamann \& 
Korista 1996). Testing the high-$Z$ results is important because of the 
possible implications for chemical enrichment and 
galaxy/structure formation in the early universe 
(Hamann \& Ferland 1993; \S5.1 below). 

Here I present a general theoretical reference for 
deriving ionizations and metal abundances in 
intrinsic absorbers, with a quantitative assessment of the 
uncertainties. I assume the column densities are accurately 
measured (or constrained) from the absorption line data. 
Accurate column densities are already available for a few bright 
QSOs, but the spectroscopy now underway with the Keck 10~m telescope 
and soon to follow from new 8~m class telescopes -- 
providing resolutions comparable to the thermal line widths -- 
will yield accurate column densities for a large number of sources. 
These improvements in the data warrant a more thorough examination 
of the theoretical uncertainties. The work presented here builds on 
previous studies such as Bergeron \& Stasi\' nska (1986), 
Chaffee \etal (1986), Steidel (1990) and Petitjean, Bergeron \& Puget (1992). 

The abundance of any metal relative to hydrogen can be derived 
from the equation,
\begin{equation}
\left[{\rm M\over H}\right] \ = \ \ 
\log\left({{N({\rm M}_i)}\over{N({\rm HI})}}\right) \ +\
\log\left({{f({\rm HI})}\over{f({\rm M}_i)}}\right) \ +\
\log\left({{{\rm H}}\over{\rm M}}\right)_{\odot}
\end{equation}
where (H/M)$_{\odot}$ is the solar abundance ratio, and $N$ and $f$ 
are  respectively the column densities and ionization fractions 
of H I and the metal M in ion state $i$. Table 1 lists several solar 
abundance ratios from Grevesse \& Anders (1989) for convenience. 
This paper will focus on the ionization corrections, 
$f$(HI)/$f$(M$_i$). The equality holds in Eqn. 1 only if the  
ionization correction applies in an average sense to the entire absorbing 
region. However, the correction factors -- in this and all previous 
studies -- derive from one-zone calculations.  
Actual absorbers can have multiple absorbing zones and a range of 
ionization states in the same system. (I will argue in \S\S4 and 5.2 
that this is, in fact, often the case.) The H I 
line(s) can form partly, or exclusively, in regions that do not 
overlap with a given M$_i$, and the column density of H I 
in the M$_i$ zone can be much less than the measured value of 
$N$(HI). Therefore, $f$(HI)/$f$(M$_i$) ratios that are 
appropriate for particular M$_i$ zones yield only lower limits to [M/H]. 

Relative abundances among the metals can also be 
potent diagnostics of the enrichment mechanism(s) and 
overall metallicity. For example, the recent reports of extremely 
high phosphorus abundances in BALQSOs, [P/C]~$\ga$~1.8 (Turnshek 1988; 
Junkkarinen \etal 1995), if confirmed, would clearly signify 
some form of non-standard enrichment (Shields 1996). 
The abundance of nitrogen relative to 
other metals, e.g. oxygen, can provide an independent measure of the 
overall metallicity because, for standard galactic enrichment, 
N/O is expected to pass from sub-solar to super-solar 
as the overall metallicity increases beyond solar 
(Wheeler, Sneden \& Truran; Hamann \& Ferland 1993). 
This behavior is due to ``secondary'' CNO processing in stellar 
envelopes, which causes 
the N abundance to grow like $\sim$$Z^2$ instead of $Z$ -- at least at 
high metallicities (see Hamann \& Ferland 1993; Vila-Costas \& Edmunds 1993). 
Relative metal abundances can be derived from the following 
analog to Eqn. 1,
\begin{equation}
\left[{a\over b}\right] \ = \ \ 
\log\left({{N({a}_i)}\over{N({b}_j)}}\right) \ +\
\log\left({{f({b}_j)}\over{f(a_i)}}\right) \ +\
\log\left({{{b}}\over{a}}\right)_{\odot}
\end{equation}
where $a_i$ and $b_j$ represent metals $a$ and $b$ in ionization 
stages $i$ and $j$, respectively. 

Abundance ratios derived from Eqns. 1 and 2 can be very uncertain 
if one is forced to invoke large (and uncertain) ionization 
corrections. For example, metallicity estimates for intrinsic absorbers 
often require large correction factors to compare H I to the 
multiply-ionized metals. If the absorbing clouds are photoionized 
by the QSO/AGN continuum, the uncertainties are due mainly to 
the unknown shape and 
intensity of the ionizing spectrum at UV through soft X-ray wavelengths. 
The recent work by Korista \etal (1996) on the BALQSO 0226--104 
demonstrates the importance of spectral shape considerations. 
Here I explore a wide range of spectral shapes consistent with 
observations. I then use the 
results to analyze (or reanalyze) some of the best published 
data and test the robustness of recent high-metallicity estimates. 
I will show that there is strong evidence for super-solar 
metallicities, even in some cases where the level of ionization is 
unknown. 

Section 2 describes the calculations and assumptions.  
Section 3 presents an array of results in the form of diagnostic diagrams.  
Section 4 applies the results to several 
well-measured \zaz\ and BAL systems. Section 5 provides a summary 
and a brief discussion of the implications.  

\section{CALCULATIONS AND ASSUMPTIONS}

I assume the absorbing gas is photoionized by the continuous radiation 
from the QSO/AGN, and I use the computer code CLOUDY (version 90.01; 
Ferland 1996) to calculate equilibrium ionization fractions under 
various conditions. 
The calculations assume static, plane parallel, constant density 
clouds illuminated on one face. The radiative intensity is 
specified by the ionization parameter $U$ -- defined as the dimensionless 
ratio of $\lambda\leq 912$~\AA\ photon to hydrogen space densities at the 
illuminated face of the clouds. The hydrogen space density is fixed at 
$n_e$~=~$10^8$~\pcc\ in all cases, although previous model experiments show 
that the ionization at a given $U$ is essentially the same for densities 
between 10$^{-2}$ and 10$^{11}$~\pcc\ (Hamann \etal 1995a). 
I assume there is no dust extinction in the absorption line regions 
or along the line of sight between the absorbers and the 
continuum source. 

\subsection{Ionizing Continuum Shapes}

Spectra of QSOs and AGNs at optical through UV wavelengths 
can be characterized by a power law ($F_{\nu}\propto \nu^{\alpha}$) 
with typical indices of $\alpha\approx -0.3$ to $-$0.7 (cf. 
Sanders \etal 1989; O'Brian, Gondhalekar \& Wilson 1989; 
Sargent, Steidel \& Boksenberg 1989; Francis \etal 1991). 
The flux must decline more sharply between the UV and 
soft X-rays in order to account for the 
observed UV to X-ray flux ratios. (In particular, 
the measured values of the power law index \aox\ 
(defined below) are typically more negative than $\alpha$ across 
the optical-UV; see Wilkes \etal 1994; Avni, Worrall \& Morgan 1995.) 
However, the poor or altogether lacking constraints on QSO/AGN spectra 
at these important energies (between roughly 10 and 500~eV, where 
the ionization thresholds of important ions appear) 
imply that there are unavoidable uncertainties in most [M/H] 
determinations (particularly those requiring a comparison between H I 
and multiply-ionized metals). 

I use four 
formulations, which I call A, B, C and PL, to explore a wide range of 
continuum shapes consistent with QSO/AGN observations. 
PL is a single power law with a freely varied slope. A, B and C  
(described mathematically in the Appendix) are segmented 
power laws with two free parameters describing the spectral shape 
between the UV and soft X-rays. The first parameter is 
\aox , which is the two-point power law index 
($F_{\nu}$~$\propto$~$\nu^{\alpha_{ox}}$) between 2500~\AA\ and 2~keV. 
The definition of \aox\ can be written,
\begin{equation}
{{F_{\nu}({\rm 2keV})}\over{F_{\nu}({\rm 2500\AA})}} = 403.3^{\alpha_{ox}}
\end{equation}
Note that this definition avoids the implicit minus sign used in most 
previous work (Zamorani \etal 1981). The second free parameter 
is the characteristic temperature $T_{c}$ (for continuum A) or 
energy $E_{c}$ (for B and C) where the 
UV spectrum gives way to a steeper decline toward the X-rays. 
Figure 1 plots the A, B and C continua for a range of these parameters. 
The figure also shows for comparison (dotted lines) the generic AGN 
spectrum derived by Mathews \& Ferland (1987, hereafter MF87), 
with an added change in slope at 0.125~eV ($\sim$10 \mum ) to 
decrease the long wavelength flux. Hereafter I will use the notation 
A(log~$T_c$,~\aox ), B(log~$E_{c}$,~\aox ) and C(log~$E_{c}$,~\aox ), 
with $T_c$ in K and $E_{c}$ in~eV, to indicate the continuum 
type (A, B, or C) and the specific parameters used. 
For example, the bold solid curves in Figure 1 represent the 
fiducial cases A(5.7,$-$1.6), B(1.4,$-$1.6) and C(1.4,$-$1.6). 

The bottom right-hand panel in Figure 1 (D) shows the MF87 and 
fiducial A, B and C continua after transmission through a column of 
highly ionized gas -- having ionization parameter $U$~=~15, 
total column density $\log~N_H$~=~22.5~\cmsq , and solar elemental 
abundances. The transmitted spectra show absorption in soft X-rays 
due mainly to bound-free transitions of O VII and O VIII at $\sim$0.8~keV. 
This type of absorption is characteristic of the so-called ``warm'' 
absorbers that appear to be common in Seyfert galaxy spectra 
(Halpern 1984; Nandra \& Pounds 1994; George \etal 1996). 
The correlated presence of X-ray warm absorbers and 
\zaz\ UV line absorbers in the majority of Seyferts and 
a few QSOs (Mathur 1994; 
Mathur \etal 1994; Mathur \etal 1995a and 1995b; 
George \etal 1995; Kriss \etal 1995; Maran \etal 1996; 
Kriss \etal 1996a and 1996b; Crenshaw 1996)
suggests that warm absorbers might lie along the line-of-sight 
between the continuum source and the UV absorption-line gas. 
This absorbing geometry also appears in some recent 
dynamical models of BAL and \zaz\ outflows (Murray \etal 1995). 
I will refer to the transmitted spectra in Fig. 1D as 
AT, BT and CT, with the shape parameters in parenthesis as before. 

\subsection{Continuum Optical Depths}

If the line-absorbing gas is optically thin in the UV through soft X-ray 
continuum, there will be a single ionization state throughout with  
the fractions $f$(M$_i$) and ratios $f$(HI)/$f$(M$_i$) for a given 
space density depending only on the shape and intensity 
of the ionizing spectrum. There is no dependence on the 
column density and very little dependence on the 
metallicity, as long 
as both remain low enough to preserve the low continuum optical 
depths\footnote{Numerical experiments show that the 
$f$(HI)/$f$(M$_i$) ratios for all ions considered here 
are the same within $\sim$0.1 dex for 
0.01~$\leq$~$Z$~$\leq$~10~\Zsun\ and $-$3.5~$\leq$~$\log U$~$\leq$~1.5 
in optically thin clouds (with the metals scaled in solar proportions). 
The only deviations larger than this are $\sim$0.3 and $\sim$0.2~dex 
depressions in the ratios involving Si III and Si IV, respectively, 
at the high $Z$ and low $U$ extremes. I therefore use $Z$~=~1~\Zsun\ 
for all calculations in the remainder of this paper.}. 
Most measurements of the column densities (\S4 below) 
in \zaz\ systems and non-Mg II BALs\footnote{I do not consider 
the minority class of BALs having low ionization lines such as Mg II 
\lam 2795, because they could have significant optical 
depths at the H I Lyman edge (Voit, Weymann \& Korista 1993).} indicate 
that the continuum optical depths are, in fact, negligibly low. 
To my knowledge, there are no observations of Lyman 
limit absorption in (established) intrinsic systems, 
and the H I column densities derived from the Lyman 
series lines are usually well below the threshold 
for optical depth unity at the Lyman edge 
(i.e. $\log~N$(HI)~$<$~17.2~\cmsq ). The column densities derived from the 
metal lines are also well below the limits for optically thick 
absorption at their bound-free edges. (For the most abundant metals, 
C, N, O and Ne, the $\tau$~$\approx$~1 limits range from 17.5~$\la$~$\log~N$(M$_i$)~$\la$~16.9 
\cmsq\ for neutral atoms to $>$18.0~\cmsq\ for highly ionized 
species; Osterbrock 1989.) Previous photoionization calculations 
applied to BALs (Weymann \etal 1985) show that the continuum optical 
depths might approach unity at the He II Lyman limit (228 \AA ). 
Numerical experiments performed here (\S3.3 below) with a variety of 
continuum shapes confirm that optical depths near unity at the 
He II edge can occur in clouds with $\log~N$(HI)~$\approx$~16.5~\cmsq , 
which is roughly the maximum H I column allowed by most observations. 
However, even in this extreme case, the marginal He II opacities 
have no significant effect on the line-of-sight-averaged ionization 
fractions. 

Therefore, unless stated otherwise, 
the calculations below assume the absorbing gas is 
optically thin in the continuum at all wavelengths. They also assume that 
the lines themselves are not significant sources of continuous 
opacity, even though Korista \etal (1996) noted that the shielding of 
continuum radiation by BALs might progressively  
soften the continuum (weaken the far-UV flux) 
farther from the QSO. I will briefly discuss 
the effects of larger continuum optical depths in \S3.3. 

\section{RESULTS}

The results in this section provide a general reference for 
deriving ionizations and metal abundances from Eqns. 1 and 2.  
Applications will be discussed in \S4 below. 
The figures and tables all use the notation C2 for 
C II, C3 for C III, C4 for C IV, etc., to save space. 

\subsection{Fully Optically Thin Clouds}

Calculations for fully optically thin clouds are widely 
applicable to intrinsic systems. Figure 2 shows 
the ionization fractions $f$(M$_i$) and $f$(HI) at various 
ionization parameters $U$ for incident spectra given by 
MF87, single power laws (PL) with 
$\alpha$~=~$-$1.0 and $-$1.5, and each of the fiducial 
continua A, B and C from Figure 1. 
The different continuum shapes can produce significantly 
different results, even for these nominal cases. For example, 
the sharp break in continuum B at energy $E_c$ (see Figure 1) 
produces smaller ionization corrections in  
C IV and N V compared to the A, C and MF87 continua, while   
the PL continua produce much larger ionization 
corrections for highly ionized elements. 

Figure 3 shows how different values of $T_c$ and \aox\ in continuum 
A affect the ionization fractions, and Figure 4 shows analogous behavior 
for different $E_c$ and \aox\ values in continuum C. 
The results for continuum B (not shown) are similar. 
One general result is that, for a given level of ionization (i.e. for fixed 
column density ratios of, for example, C III/C IV or N III/N V), 
the ionization corrections are larger 
for larger values of $T_c$ and $E_c$. This is because the increased far-UV 
flux shifts the $f$(M$_i$) curves toward lower $U$ where $f$(HI) 
is larger. Therefore, adopting continuum shapes with large 
$T_c$ or $E_c$ leads to larger derived metallicities.  
For example, a given observed ratio of $N$(C III)/$N$(C IV)~=~1 column 
densities yields $\sim$0.8 dex larger [C/H] for the largest compared to 
the smallest $T_c$ and $E_c$ values in Figures 3a and 4a. 
Larger (less negative) values of \aox\ also favor 
larger ionization corrections and thus larger metal abundances. 

If the measured column densities provide no constraints on the ionization, 
or if they indicate a range of ionizations, it might be reasonable to 
assume that each line forms near the peak in its $f$(M$_i$) 
curve (cf. Hamann \etal 1995a). This assumption, in any case, yields 
conservatively low estimates of [M/H] for moderately ionized 
metals like C IV and N V because their ionization 
corrections are nearly minimized at the $f$(M$_i$) peaks (see below). 
Figures 5 and 6 plot the normalized corrections, 
log($f$(HI)/$f$(M$_i$))~+~log(H/M)$_{\odot}$, evaluated at 
the $f$(M$_i$) peaks for a wide range of continuum A and C shapes. 
As expected, the correction factors 
for highly ionized species are very sensitive to the continuum 
shape. The corrections are larger for large 
values of $T_c$, $E_c$ and \aox , leading to larger derived 
[M/H] in these cases.  

Figure 7 shows that the correction factors for a given continuum shape 
all reach minimum values at finite $U$. 
These minima occur near (but at slightly 
larger $U$ than) the peaks in the $f$(M$_i$) curves (compare Figs. 2 
and 7). Figures 8 and 9 plot the normalized minimum ionization corrections 
(the quantity log($f$(HI)/$f$(M$_i$))~+~log(M/H)$_{\odot}$  
evaluated at the minimum of $f$(HI)/$f$(M$_i$)) for 
various continuum A and C shapes. These minimum correction factors 
can be used with Eqn. 1 to place firm lower limits on [M/H]. 

Figure 10 shows the normalized ionization corrections needed to 
derive the relative metal abundances N/C, N/O and P/C from several ionic 
ratios involving these elements. The quantity plotted for each ionic ratio 
corresponds to the last two terms in Eqn. 2. The results utilize 
the same calculations as Figs. 2 -4, for optically thin 
clouds photoionized by different A and C continuum shapes. 

Ideally, one would derive relative metal abundances from ions with 
similar ionization potentials so that (1) the measured lines form as much 
as possible in the same gas and (2) the ionization correction is 
small and insensitive to the uncertain spectral shape and intensity. 
Figure 10 shows that, among the more readily measured column densities, 
N III/C III\ provides the most reliable 
indicator of the relative nitrogen abundance. The figure also shows 
that there is a minimum ionization correction for P~V/C IV, which
can be used to place firm lower limits on the P/C abundance.  

\subsection{Fully Optically Thin Clouds Shielded by a Warm Absorber}

If an X-ray warm absorber lies between the line-absorbing 
clouds and the QSO/AGN continuum source, the incident spectra 
could resemble the curves plotted in Figure 1D. 
The left-hand panel in Figure 11 shows the ionization fractions 
in optically thin clouds irradiated by the transmitted continuum 
CT(1.4,$-$1.6). (The ionization state of the warm absorber and 
its possible contribution to the UV absorption lines are discussed in 
\S\S3.3 and 5.2 below.) The intervening 
warm absorber depresses the flux incident on the UV-line clouds 
at soft X-ray wavelengths and thus alters the ionization fractions 
for highly ionized species such as Mg X, Ne VIII, O VIII, O VII 
and, to a lesser extent, O VI. However, the intervening 
warm absorber does not 
substantially alter the $f$(HI)/$f$(M$_i$) ratios for lower ionization 
metals, including C IV and N V. Therefore, one could still use the results 
from \S3.1 for most abundance determinations. 

\subsection{Marginally Optically Thick/Warm Absorber Clouds}

Another possibility is that the UV lines form within the  
X-ray warm absorber itself (Mathur \etal 1994; Mathur 1994; Mathur \etal 
1995a and 1995b). Column densities derived from observed UV absorption 
lines generally require low continuum optical depths at the ionization 
thresholds of H I and the low- to moderate-ionization metals 
(up to at least the C IV and N V; \S2.2 and \S4 below). But the data 
do not directly constrain the opacities due to He I and He II, and 
they do not rule out significant opacities in the soft X-rays due to 
highly ionized metals. (The measured low Ne VIII 
column density in the QSO UM~675 is a notable exception; see Hamann \etal 
(1995a) and \S5.2 below.) 
If the line-forming gas has significant continuum opacities anywhere across 
the UV through soft X-ray spectrum, there can result a range of ionization 
states in one-zone environments with the amount of low-ionization 
material depending on the total column density. 
This one-zone, multi-state situation can yield very different ionization 
corrections from the fully optically thin cases described above. 

The right-hand side of Figure 11 shows the mean ionization fractions 
for clouds having roughly the maximum continuum optical depths allowed 
by observations (\S2.2). 
The specific parameters for this calculation are an optical depth 
at the H I Lyman limit of $\tau_{LL}$~$\leq$~0.2 (or equivalently 
$\log~N$(HI)~$\leq$~16.5~\cmsq ), 
a total hydrogen column density of $\log~N_H$~$\leq$~22.0~\cmsq , 
and solar abundances.  
The upper limit on $\tau_{LL}$ keeps the clouds from becoming 
``cold'' absorbers at low $U$, which would contradict the low 
column densities measured for H I and the low- to moderate-ionization 
metals (\S2.2). The $\tau_{LL}$~$\leq$~0.2 requirement is 
enforced by truncating the total column density at 
$\log~N_H$~$<$~22.0~\cmsq\ if necessary. These low-$U$ clouds 
are not warm absorbers. They have low optical depths at all 
wavelengths, except at the He II Lyman limit where they reach 
optical depths up to unity. At high ionizations (log~$U$~$\ga$~0.0), 
the full column 
of $\log~N_H$~=~22.0~\cmsq\ is realized and the clouds become warm 
absorbers with considerable soft X-ray opacity. These high-$U$ calculations  
encompass the warm absorber models described by Mathur and collaborators. 

The ionization fractions in Figure 11 differ from the fully 
optically thin results only for the highly ionized metals in high-$U$ 
clouds, i.e. in the warm absorber regime (compare to Fig. 2b). 
The H I and low- to moderately-ionized metals are unaffected 
by the finite column densities because the gas is still required 
to be optically thin at most UV through far-UV wavelengths. 
Therefore, as in \S3.2, one can still use the fully optically 
thin results to derive abundances from these ions.  

\subsection{Two Minor Cautions}

Before applying the results above to actual data, I note 
two minor points of caution. First, the ionization state of the gas 
could be out of equilibrium with the radiation field if the 
intensities vary significantly on timescales comparable to, 
or shorter than, the recombination times. 
Krolik \& Kriss (1995) showed that non-equilibrium effects can 
occur in low density gas, $n_e$~$\la$~$10^6$ \pcc , 
near rapidly varying Seyfert 1 galaxies. 
I have ignored this complication 
because the spectral variations in luminous QSOs should have much 
lower amplitudes than the Seyferts (Christiani \etal 1996), and because  
the densities in at least the BAL regions should be 
higher than $\sim$$10^6$ \pcc\ (Turnshek 1988; Weymann \etal 
1985; Barlow \etal 1992; Barlow 1993). Furthermore, 
if the ionization does lag behind changes in the continuum flux, 
the ionization state at any epoch represents an average over the 
recent spectral variations (see Krolik \& Kriss 1995 for discussion). 
Therefore the calculations here can be interpreted 
as averages over the spectral and ionization fluctuations.

Second, some of the highest ionization clouds depicted in the 
figures above are unstable to changes in the gas pressure. 
Instabilities can occur for gas temperatures 
$T_e$~$>$~10$^5$~K (cf. Field 1965; Gehrels \& Williams 1993), 
but the ranges in $U$ for instability depend 
on the shape of the ionizing spectrum. 
For an incident MF87 continuum, the gas is unstable for 
1.5~$\la$~$\log~U$~$\la$~1.8 
and 2.2~$\la$~$\log~U$~$\la$~2.5 (Hamann \etal 1995b; 
see also Kallman \& Mushotzky 1985; 
Netzer 1990; Marshall \etal 1993; and Krolik \& Kriss 1995). 
The temperatures and ionizations 
inferred generally for BAL and \zaz\ regions are well below 
these unstable regimes (also Netzer 1996). 

\section{APPLICATIONS: \ METAL ABUNDANCES IN QSOs}

Here I use the calculations for optically thin, unshielded clouds 
(\S3.1) to estimate [M/H] from the best published column densities 
for BAL and \zaz\ systems. I consider a range of 
continuum shapes consistent with the observations of each QSO and 
use those ranges to derive theoretical uncertainties. 
Table 2 lists the sources, their emission redshifts, the type of 
absorption system (BAL or \zaz ), and the adopted spectral 
parameters for continua A and C. Usually no observational constraints 
are available on the far-UV continua. I therefore adopt 
log~$T_c$~=~5.7$^{+0.3}_{-0.5}$ for continuum A and 
log~$E_c$~=~1.4$^{+0.5}_{-0.3}$ for continuum C, unless noted 
otherwise below. Values of $T_c$ and $E_c$ outside of these ranges 
are unlikely because they would violate most QSO/AGN observations 
at UV and X-ray wavelengths. (The resulting spectra would either 
decline to steeply toward higher energies across the optical-UV [small 
$T_c$ and $E_c$] or rise too steeply toward lower energies 
in X-rays, producing too-large soft X-ray excesses [large $T_c$ and $E_c$]; 
see \S2.1 and the Appendix for references to observations). 
If no measurements of \aox\ are available, 
I use the redshifts and optical magnitudes (from Hewitt \& 
Burbidge 1993) to estimate \aox\ from the luminosity--\aox\ 
relation given by Wilkes \etal (1994, their equations 6 and 12; 
see also Avni \etal 1995). In those cases, I adopt 
uncertainties of $\pm$0.2 for \aox\ based on the measured 
scatter among QSOs of a given luminosity (Avni \etal 1995). 

The abundance results are summarized in Table 3 
and in the notes on individual systems below. 
For each system, Table 3 lists the absorption redshift, 
the velocity ($\Delta v$ in \kms ) relative to the emission redshift 
(for \zaz\ systems only), the measured 
column densities in H I (log $N$(HI) in \cmsq ) and each metal ion 
(log~$N$), and three pairs of [M/H] and $U$ 
results. The first two [M/H]--$U$ pairs are the best estimates 
of these quantities for clouds ionized by 
continua A and C respectively, with $U$ defined whenever possible by column 
density ratios such as C II/C III/C IV, N III/N V, 
Al II/Al III or Si II/Si III/Si IV. 
When no firm constraints on $U$ are available, or when different constraints 
yield significantly different $U$ values, I assume the lines 
form at the peaks in their $f$(M$_i$) curves and derive 
[M/H] from Figures 5 and 6. This assumption is 
invoked for the majority of systems in the table, as indicated by a `p' 
in the third column. It yields conservatively low [M/H] 
values for moderately ionized metals like C IV and N V because the 
ionization corrections are near their minima at the $f$(M$_i$) peaks 
(e.g. compare Figs. 5 and 8 in \S3.1). Even in cases where 
$U$ is well-constrained by one or more column density ratio, the [M/H] 
in Table 3 are probably often underestimates, particularly for C IV, N V and 
more highly ionized species, because only some of the measure H I will 
reside with each metal ion in multi-phase absorbers (\S1 and \S5.1 below). 

The last [M/H]--$U$ pair in Table 3 indicates the absolute minimum 
metal abundance, [M/H]$_{min}$, and corresponding $U_{min}$ 
for ionization by continuum C (from Figure 9). The results for 
continuum A (not listed) are similar. These minimum abundances 
apply for {\it any} single- or multi-phase medium photoionized by 
continuum C (see \S3.1). 
The uncertainties listed for all of the [M/H]--$U$ pairs show the range 
of values allowed by the range of continuum shapes in Table 2. 
The uncertainties in the measured column densities (when 
available from the literature) are also listed in Table 3 (in column 1), 
but they are not included in the [M/H]--$U$ estimates. Please 
see the references cited below for more information on the measurement 
uncertainties. 

\subsection{Notes on Individual \zaz\ Systems}

\subsubsection{Q 0000$-$263}

Bechtold \etal (1994) derive 
\aox\ $\approx$ $-$1.85 for this QSO 
based on UV and X-ray observations separated 
by three months. The figures in Bechtold \etal reveal 
an unusually steep UV spectrum with $\alpha_{uv}$ $\sim$ $-$1.6 to $-$2.0 
between roughly $\sim$2400 and $\sim$900~\AA\ in the rest frame. I 
therefore adopt low values of $T_c$ and $E_c$, with uncertainties that 
encompass the wide range of shapes still permitted in the far-UV. 
The column densities in Table 3 are from Savaglio \etal (1994). 
The ratio of Si III/Si IV columns defines the ionization 
parameter in the two highest redshift systems. There is strong evidence 
for [M/H]~$\ga$~0.0 in three of the eleven systems: 
$z_a$~=~4.1270, 4.1323 and 4.1334. 

\subsubsection{UM 675}

The UV continuum is observed 
to turn down sharply blueward of the \Lya\ emission line 
($\sim$1200~\AA ; Beaver \etal 1991). 
However, it is not clear that this spectral shape is 
intrinsic to the QSO or that it continues beyond the shortest 
observed rest-frame wavelength of $\sim$520~\AA . 
Lyons \etal (1996) show that 
the intrinsic UV continuum can, in fact, be interpreted as a simple 
power law with $\alpha_{uv}$ $\sim$ $-$1  
modified by many blended, intervening \Lya\ absorption lines 
(the so-called Lyman ``valley''; M\o ller \& Jakobsen 1990). 
Here I adopt somewhat low values of $T_c$ 
and $E_c$ consistent with the observed spectrum, 
but with uncertainties that encompass the usual range of 
possible intrinsic shapes. The column densities 
(from Hamann \etal 1995a) require a range of ionizations (i.e. 
more than one $U$ value). It is clear from Table 3 that the 
H I forms mostly in the lower ionization gas e.g. with C III, C IV and 
N III. As a result, the [M/H] values derived assuming each ion 
forms at the peak in its $f$(M$_i$) curve are greatly 
underestimated for highly ionized species such as O VI and Ne VIII. 

The \zaz\ system in this QSO is of particular interest because its 
time-variability places the absorbers within a few hundred pc of the 
continuum source (Hamann \etal 1995a). In agreement with Hamann \etal 
(1995a), Table 3 provides strong evidence for [C/H]~$\ga$~0.0 
and [N/C]~$\ga$~0.8. 
Note that the revised column densities derived for this source 
by Hamann \etal (1997b), 
which make an uncertain correction for the partial coverage of the 
background light source, would imply larger M/H ratios by factors of $\sim$2.

\subsubsection{PKS 0424$-$131} 

Wilkes \etal (1994) 
derive \aox\ $\approx$ $-$1.60 based on non-simultaneous UV and X-ray 
observations. I assign an uncertainty of $\pm$0.1 due to the 
possible variability and measurement/fitting uncertainties. 
The column densities are from Petitjean 
\etal (1994), who take explicit account of the partial coverage of the 
background emission source(s) when necessary. 

$z_a$~=~2.100. The ratios of C II/C IV and Si II/Si III/Si IV 
column densities suggest that there is a range of ionizations in this 
system. The [M/H] values for the low-ionization metals  
indicate metallicities 0.0~$\la$~[M/H]~$\la$~1.0 for 
the maximum H I column listed in Table 3. Line fits by Petitjean 
\etal indicate that the H I column could actually be a factor of 
$\sim$10 lower, which would imply commensurately higher metallicities. 

$z_a$~=~2.133. The C IV and N V 
column densities apply to the sum of two line 
components fit by Petitjean \etal (1994). The H I column might arise 
from either or both of the two systems. 

$z_a$~=~2.173. The ionization parameter for this systems was estimated 
from the Si III/Si IV column density ratio. 

\subsubsection{PKS 0450$-$132} 

York \etal (1991) give a $B$ magnitude 
of 17.5, which yields \aox\ $\approx$ $-$1.7 following 
Wilkes \etal (1994).  
The column densities are from Petitjean \etal (1994), where again 
partial coverage is taken into account when necessary. The two systems 
listed in Table 3 are just single components in multi-component blends. 
The H I columns are upper limits based on line fitting. 

$z_a$~=~2.1050. The ionization parameter adopted for this system 
is an average of two similar values derived from the C II/C IV and 
Si II/Si IV column density ratios. The results in Table 3 indicate that most 
the H I forms with the low ionization metals. The [M/H] values for those 
ions imply 0.0~$\la$~[M/H]~$\la$~0.5 for the maximum H I 
column density. Line fitting 
by Petitjean \etal (1994) shows that the H I column could actually 
be a few times lower, which would imply commensurately larger metallicities. 

$z_a$~=~2.2302. The ionization parameter follows from the Si III/Si IV 
column density ratio. The [M/H] values for the 
Si ions imply metallicities of order solar 
or higher, assuming the maximum H I column density. 
Petitjean \etal (1994) indicate that the actual H I column could be a 
factor of $\sim$10 lower, which would again imply larger [M/H].  

\subsubsection{HS 1946+769} 

Bechtold \etal (1994) derive \aox\ $\approx$ $-$1.93 based on UV and 
X-ray observations made three months apart. The UV continuum for 
$\lambda_{rest}$ $>$ 740~\AA\ is discussed 
by Kuhn \etal (1995), but the far-UV continuum is not 
constrained. The column densities are from Tripp \etal (1996). 
Those listed for $z_a$~=~3.0496 are Tripp {\it et al.}'s best estimates 
for one component of a severely blended, multi-component 
system (from their Table 5). The upper limit on the H I column comes 
from the lack of Lyman limit absorption ($\tau_{LL}$~$<$~0.2) associated 
with this system (see also Fan \& Tytler 1994). 
I estimate $U$ for the $z_a$~=~3.0496 system from the ratio of 
Al II/Al III column densities. The resulting ionization is consistent 
with the measured Si III/Si IV ratio, and with the limits on 
Si II/Si IV, C II/C IV and N V/C IV (for reasonable N/C abundances). 
Note that all of the 
[M/H] values listed for this system are only lower limits, because 
the H I column density is an upper limit and in some cases 
the metal columns are lower limits (e.g. C IV, Si III and Si IV). 
The results are consistent with Tripp \etal (1996). 

\subsubsection{Q 2116$-$358} 
  
The column densities are from Wampler \etal (1993). 
Table 3 does not include the associated system at 
$z_a$~=~2.318434 because its H I column density is 
poorly known. Wampler \etal estimate $\log~N$(HI)~$\la$~17.6 \cmsq , 
which implies $\tau_{LL}$~$\la$~2.5 at the Lyman limit. 
For the other four systems near $z_a$~$\approx$~2.318 (Table 3), 
the column density ratios of different ions of C and Si 
imply a range of $U$ values -- sometimes spanning more than 
an order of magnitude in the same system. I therefore assume all of 
the ions form at the peaks in their $f$(M$_i$) curves. 
Wampler \etal describe evidence for partial coverage in several of the 
systems, but it is not clear that they took that into account in 
deriving the column densities. 
If the column densities are accurate, there is strong evidence 
for [M/H]~$\ga$~0.0 in all nine systems and for [M/H]~$\ga$~1.0 
in a few. 

It should be noted that the (probably) large H I column in the system 
at $z_a$~=~2.318434 could significantly lower the Lyman continuum flux 
incident upon any clouds farther from the QSO. I did not take this 
into account in Table 3, but the net effect on 
more distant clouds would be to shift the $f$(M$_i$) curves to the left 
in Figures 2 -4 and thus {\it increase} the $f$(HI)/$f$(M$_i$) 
ratios and the derived metal abundances by factors up to $\sim$10. 

\subsubsection{PG 2302+029}

The lines in this unusual system are dominated by a single 
broad component with FWHM~$\sim$~3000 
to 5000 \kms\ and centroids displaced by $-$56,000 \kms\ from the 
emission redshift (Jannuzi \etal 1996). The absorbers could be 
ultra-high velocity ejecta from the QSO, possibly related to the 
BAL phenomenon, or intervening gas distributed on cosmologically 
significant scales, perhaps in a foreground cluster of galaxies. 
I include this system here assuming that the clouds, whatever 
their location, are optically thin in the 
continuum and in photoionization equilibrium with a QSO spectrum. 

The near-UV spectrum of the QSO is somewhat steep, 
with $\alpha\approx -1.25$ for $\lambda_{rest} > 1216$ \AA\ 
(O'Brian \etal 1988). I therefore adopt low fiducial values 
of $T_c$ and $E_c$. 
The column densities in Table 3 follow from the equivalent widths 
in Jannuzi \etal (1996) assuming the lines have negligible 
optical depths. The upper limit on the H I column is based on 
my own estimate of $\la$4 \AA\ for the \Lya\ equivalent 
width. The derived minimum C/H ratio is not much below solar, 
consistent with absorption near the QSO. However, Jannuzi \etal point out 
that the measured column densities are consistent with much lower 
metallicities if the gas is collisionally ionized (compare, for example, 
Figs. 3 and 5 in Hamann \etal 1995a). Therefore, the metallicity 
results in Table 3 cannot be used to determine the intrinsic {\it vs.} 
intervening origin of the lines. 

\subsection{Notes on Individual BAL Systems}

\subsubsection{Q 0226$-$104} 

Constraints on the 
continuum shape are discussed by Korista \etal (1996). 
The spectrum steepens sharply at rest wavelengths below $\sim$1200~\AA , 
but it is not clear that this shape is intrinsic to the QSO or that it 
continues across the entire far-UV spectrum. I therefore 
adopt slightly low values 
of $T_c$ and $E_c$ to be consistent with the observed spectral shape, but 
with uncertainties that encompass the usual range of shapes. 
I adopt Korista {\it et al.}'s estimate of \aox\ $\approx$ $-$1.7 
from the relations in Wilkes \etal (1994). 

The wide range of column 
densities measured for this source -- from combined ground-based and 
Hubble Space Telescope observations (Korista \etal 1992 and 1996) -- 
provide the best constraints available on the ionization and 
abundances in a BALQSO. 
Korista \etal (1996) note that the column densities measured from the 
ground (H I, C IV, N V and Si IV) are typical of BALQSOs, and therefore 
the derived abundances and ionizations might also be typical. The results 
in Table 3 are consistent with the previous studies of this source 
(Korista \etal 1996; Turnshek \etal 1996), indicating a range of 
ionizations with super-solar metallicities and N relatively more 
enhanced.

\subsubsection{Q1246$-$057 and RS 23} 

The column densities are from Junkkarinen, 
Burbidge \& Smith (1987). These two sources were chosen from Junkkarinen 
{\it et al.'s} sample of six BALQSOs because their absorption troughs are 
less confused with the emission lines -- making the column densities 
more reliable. Super-solar metallicities are required for both objects, 
even though the level of ionization is unknown. 

\subsubsection{Mean BALQSO} 

The column densities were measured by Hamann \etal (1993) 
from the mean spectrum of 36 BALQSOs having weak or absent low 
ionization lines (Weymann \etal 1991). Super-solar metallicities are 
again required for this average BALQSO. 

\section{SUMMARY AND DISCUSSION}

The figures in \S 3 provide a general reference for estimating the 
ionization and metal abundances in clouds that are photoionized by 
a QSO/AGN. The poor or altogether lacking constraints on the UV through 
soft X-ray spectra of QSOs/AGNs imply that there are unavoidable 
uncertainties in the ionization corrections and derived abundances. 
Nonetheless, the calculations applied to the best measured column 
densities for QSO intrinsic absorbers 
(from the literature) provide strong evidence for super-solar 
metallicities in BALs and at least some \zaz\ systems. 
Even when the continuum shape is uncertain and 
the level of ionization is unknown (for example, when only H I and 
C IV lines are detected), firm lower limits on the ionization 
corrections imply [M/H]~$>$~0.0 for typical BALs and some \zaz\ systems. 
Conversely, none 
of the 32 systems considered here require $Z$~$<$~\Zsun . 
These results support the previous claims for typically high 
metallicities in QSO intrinsic absorbers (\S1). They also 
support the independent evidence for 
high metallicities in QSOs based on the broad emission lines 
(Hamann \& Ferland 1992 and 1993; Hamann \& Korista 1996; 
Ferland \etal 1996). 

The measured column densities for intrinsic absorbers indicate 
that the absorbing gas is usually optically thin in 
the H I Lyman continuum out to energies 
beyond at least the C IV and N V ionization edges (at 64 eV 
and 98 eV, respectively). 
Column density ratios such as C II/C III/C IV, N III/N V and 
Si II/Si III/Si IV often imply a range of 
ionization states in the same system. Neighboring \zaz\ systems 
(having similar redshifts and often blended spectroscopically) 
also typically have different ionizations 
or abundances. If the abundances are similar in neighboring systems, 
their ionization parameters must sometimes differ by more than an order 
of magnitude. To achieve different ionization states in optically 
thin gas, the absorbers (at the same 
or similar redshifts) must span a significant range of densities or 
distances from the ionizing QSO. 

\subsection{High Metallicities at High Redshift}

The general result for solar or higher metallicities in QSOs 
is consistent with normal galaxy evolution. 
Vigorous star formation in the cores of massive galaxies 
can produce gas-phase metallicities well above 
solar within a few billion years of the initial collapse 
(cf. Arimoto \& Yoshii 1987; K\" oppen \& Arimoto 1990). 
High metal abundances are a signature of {\it massive} 
galaxies because only they can retain their gas long enough 
against the building thermal pressures from supernova explosions. 
The enriched nuclear gas might ultimately be ejected from the galaxy 
or consumed by the black hole, 
but the evidence for evolution to high $Z$ remains today in the stars.  
For example, the stars in the central bulge 
of our own Galaxy have a broad distribution of metallicities, with a mean 
of $\sim$1~\Zsun\ and a maximum of $\sim$3~\Zsun\ (McWilliam \& Rich 
1994; Rich 1995 -- private comm.). If the gas is well-mixed during the 
evolution, its final metallicity will equal that of the most metal 
rich stars (i.e. $\sim$3~\Zsun ). Simple closed-box (or, equivalently, 
rapid infall) models 
of the chemical enrichment (Tinsley 1980; Searle \& Zinn 1972), 
which fit the stellar $Z$-distribution in the bulge very well (Rich 1990), 
also predict final gas-phase metallicities of $\sim$3~\Zsun\ 
(assuming $\sim$95\% conversion of the original gas into stars). 
The same calculations applied to the higher mean 
stellar metallicities of $\sim$2 to 3~\Zsun\ observed 
in the cores of elliptical galaxies and some nearby 
spirals (cf. Bica 1988; Bica, Arimoto \& Alloin 1988; 
Bica, Alloin \& Schmidt 1990; Gorgas, Efstathiou \& Arag\' on Salamanca 
1990; Worthey, Faber \& Jes\' us Gonzalez 1992; Jablonka, Alloin \& Bica 
1992; Kennicutt \& Garnett 1996) imply even higher gas-phase abundances  
of $\sim$6 to 9~\Zsun . These simple estimates are supported by 
more sophisticated simulations of the enrichment 
(cf. Hamann \& Ferland 1993 
and references therein). Therefore, super-solar metallicities can be 
{\it expected} in QSO environments as long as (1) the gas 
is processed in the cores of massive ($\ga$10$^{11}$~\Msun ) 
galaxies (or at least in dense condensations that become the cores 
of massive galaxies), and (2) the enrichment timescales are shorter than  
the time needed for QSOs to ``turn on'' or become observable 
(e.g. less than a few Gyr for QSOs at redshifts $\ga$4; 
Hamann \& Ferland 1993). 

In addition to this normal stellar enrichment, other processes peculiar 
to black-hole/accretion-disk environments might contribute to, or 
even dominate, QSO abundances (Artymowicz, Lin \& Wampler 1993; Shields 
1996). Metallicities above $\sim$10~\Zsun\ are problematic for 
standard enrichment models unless the initial mass functions 
strongly favor massive stars. The results in \S4 indicate 
that $Z \ga 10$~\Zsun\ occurs in typical BAL and some 
\zaz\ regions. Possibly an additional source of metals is required, 
or possibly the gas does not derive from a well-mixed interstellar 
medium. The extremely high phosphorus abundances of [P/C]~$\ga$~1.8   
reported for a few BALQSOs (Turnshek 1988; Junkkarinen \etal 1995) 
have led to speculation that BALs form in a special environment, 
such as dwarf novae ejecta (Shields 1996). The high [P/C] values are 
in any case not compatible with the normal stellar enrichment of a 
well-mixed interstellar medium. These results should be 
tested in other objects. Presently, there is no evidence for large P/C 
ratios in \zaz\ absorbers or in the broad emission-line regions. 

\subsection{Intrinsic UV Absorbers as X-ray Warm Absorbers?}

There is no evidence from the data discussed here that the 
line-absorbing clouds also produce continuous X-ray absorption. 
Bound-free absorption in soft X-rays due to highly ionized metals, 
mainly O VII and O VIII, occurs in many Seyfert 
galaxies and at least some QSOs (e.g. Nandra \& Pounds 1994; 
George \etal 1996 and references therein). 
Mathur (1994), Mathur \etal (1994), Mathur \etal (1995a) and 
Green \& Mathur (1996) argued that BALs 
and \zaz\ lines form in the same gas as the X-ray warm absorbers. 
In their models, the ions producing the common UV lines 
are minor constituents in highly ionized gas with 
log~$U$~$\ga$~$-$0.5. These one-zone models predict strong 
trends for larger column densities in more highly ionized species 
(see also Netzer 1996 and Shields \& Hamann 1996). For example, a modest 
ionization of $\log~U$~=~0.0 produces logarithmic 
column density ratios of C II/C III/C IV, N III/N V, O III/O VI and  
Si II/Si III/Si IV equal to $-$5.8/$-$2.1/0.0, $-$3.3, $-$3.9, and 
$<$$-$5.2/$-$2.1/0.0, respectively, in a nominal warm absorber 
(\S3.3, Fig. 11). If the metals have solar relative abundances, 
the same calculation predicts logarithmic column density ratios such 
as Mg II/C IV~$<$~$-$9.1, Si IV/C IV~=~$-$3.9, N V/C IV~=~+0.3, 
O VI/C IV~=~+1.7, Ne VIII/C IV~=~+0.1. 

Although one-zone models are consistent with the combined UV and 
X-ray observations of several Seyfert galaxies and QSOs (see papers by 
Mathur \etal above; also Shields \& Hamann 1996), the more reliable 
(e.g. higher spectral resolution) absorption-line 
data discussed here do not support a single-zone absorber. 
In particular, the strongs trends for much 
larger column densities in more highly ionized species are not present 
in the data. The measured column densities often 
indicate a range of ionizations, with most of the lines forming at 
$\log U$~$\la$~$-$1.0. The significant presence of singly-ionized 
metals alone cannot be reconciled with 
a highly ionized, one-zone absorber when the gas is {\it known} 
to be optically thin in the H I Lyman continuum. 
In the one case where warm absorber-like ionizations are known 
to be present (via Ne VIII~\lam 770,780 in the \zaz\ system of UM~675), 
the column density is roughly two orders of magnitude too small 
to be a warm absorber (Hamann \etal 1995a). The ensemble data for 
that system (Table 3) indicate 
that the gas is optically thin in the continuum at {\it all} UV through 
X-ray wavelengths. 

Nonetheless, the detections of strong O VI~\lam 1032,1038 and 
especially NeVIII~\lam 770,780 are significant. Few observations 
so far have included these lines, but O VI, at least, appears to 
be common. There might often be {\it components} of 
UV absorbing regions that have high, warm absorber-like ionizations. 
Perhaps sometimes, and perhaps more often in Seyfert galaxies, the 
highly ionized components {\it are} warm absorbers. 
More efforts to combine X-ray observations with accurate UV-derived 
column densities across a range of ionizations, including O VI and 
Ne VIII, are needed to explore the relationship between the 
two absorbers further. 

\acknowledgments
I am grateful to G. J. Ferland and K. T. 
Korista for their generous help with Ferland's CLOUDY software. 
Thanks to V. Junkkarinen and G. Shields for useful discussions 
and J. C. Shields, K. T. Korista and the referee P. M\o ller 
for helpful comments on this manuscript. This work was supported by NASA 
grants and AR-5292.02-93B, NAG 5-1630 and NAG 5-3234.  

\bigskip\bigskip\bigskip
\centerline{\bf APPENDIX. \ Continuum Shape Details}
\smallskip

The ionizing continua A, B and C are all consistent with generic 
QSO/AGN observations. They each portray a range of 
shapes at the crucial UV through soft X-ray wavelengths. 
Continuum A (Figure 1A) has the form, 
\begin{equation}
F_{\nu} \propto 
\nu^{\alpha_{uv}}e^{-{{kT_{ir}}\over{h\nu}}}e^{-{{h\nu}\over{kT_c}}} 
+ \Gamma\nu^{\alpha_{x}}e^{-{{kT_{c2}}\over{h\nu}}}e^{-{{h\nu}\over{kT_x}}}
\end{equation}
which is the sum of two power laws, each having 
exponential cutoffs at high and low energies. I adopt a power law 
index in the near-infrared through UV, $\alpha_{uv}$~=~$-$0.5, 
consistent with the wealth of QSO/AGN 
observations cited in \S2.1. The infrared cutoff has a fixed characteristic 
temperature of $T_{ir}$ $\approx$ 300~K (equivalent to 
$\sim$10 \mum ), which eliminates free-free heating at high densities 
(Ferland \etal 1992). The important high energy UV cutoff, $T_c$, is 
freely varied; the range 5.0 $\leq$ $\log~T_c$ $\leq$ 6.0~K is shown in 
Figure 1A. The relative strengths of the UV and X-ray continua 
are set by the constant $\Gamma$ according to the definition of \aox\ 
(Eqn. 3). The X-ray continuum has a power law index of 
$\alpha_x$~=~$-$1.0. This slope is typical of radio-quiet 
QSOs (Wilkes \& Elvis 1987) and Seyfert galaxies (Nandra \& Pounds 
1994) at energies 2--10~keV. 
The energetically more important soft X-ray spectra 
are usually steeper than this below $\sim$0.5 to 1~keV 
(Wilkes \& Elvis 1987; Turner \& Pounds 1989; 
Wilkes \etal 1994; Walter \etal 1994; 
Gondhalekar \etal 1994; Fiore \etal 1994; Laor \etal 1994; 
Marshall, Fruscione \& Carone 1995). Therefore, 
the choice of $\alpha_x$~=~$-$1.0 is probably also a good compromise 
for radio-loud QSOs, even though their 2--10~keV slopes are typically 
closer to $-$0.5 (Wilkes \& Elvis 1987). 
The X-ray power law is suppressed at low energies by an 
exponential decline characterized by the temperature 
$T_{c2}$~=~$T_c$/6. At the highest energies, the X-ray flux declines with a 
characteristic temperature of $\log T_x$~=~9.0~K. These cutoff 
temperatures are 
simply a mathematical convenience for producing a smooth, well behaved 
continuum overall. 

Continuum B (Figure 1B) is a segmented power law 
with changes in the index at energies 0.125~eV, $E_{c}$, and $E_{cx}$. 
The power law indices between these energies are,
\begin{eqnarray}
\alpha_{ir} & = & +2.5 {\hskip 1.0cm} {\rm for}\  
0 < h\nu  \leq 0.125 {\rm~eV} \nonumber\\
\alpha_{uv} & = & -0.5 {\hskip 1.0cm} {\rm for}\  0.125 {\rm~eV} < 
h\nu  \leq E_{c} \nonumber\\
\alpha_{c} & = & -3.5 {\hskip 1.0cm} {\rm for}\ E_{c} < h\nu  \leq E_{cx} \\
\alpha_{x} & = & -1.0 {\hskip 1.0cm} {\rm for}\  E_{cx} < h\nu  < \infty
\nonumber
\end{eqnarray}
The freely varied parameters are \aox\ and $E_{c}$, with the value of 
$E_{cx}$ determined from the definition of \aox\ in Eqn. 3. 

Continuum C (Figure 1C) is a broken power law like continuum B, but with 
$E_{cx}$~=~0.7~keV fixed in all cases. This break in the spectral 
slope at 0.7~keV is consistent with observations 
of soft X-ray ``excesses" 
(Wilkes \& Elvis 1987; Wilkes \etal 1994; Walter \etal 1994; 
Fiore \etal 1994; Laor \etal 1994; Marshall \etal 1995). 
Continuum C differs from B in that it always includes a 
soft X-ray excess, and the slope $\alpha_c$ between $E_c$ and $E_{cx}$ 
depends on the adopted values of \aox\ and $E_c$. 
For example in Figure 1C, $\alpha_c$ varies from $-$1.72 for 
C(0.9,$-$1.6) to $-$3.02 for C(1.9,$-$1.6). 

\bigskip\bigskip
\begin{center}
{\bf REFERENCES}
\end{center}
\parskip=0pt
\leftskip=0.5in
\parindent=-0.5in


Anderson, S. F., Weymann, R. J., Foltz, C. B., \& Chaffee Jr., F. H. 
1987, AJ, 94, 278

Arimoto, N. \& Yoshii, Y. 1987, A\&A, 173, 23

Artymowicz, P., Lin, D. N. C., \& Wampler, E. J. 1993, ApJ, 409, 592

Avni, Y., Worrall, D. M., \& Morgan, W. A. 1995, ApJ, 454, 673

Baldwin, J. A., Ferland, G. J., Korista, K. T., Carswell, R. F., 
Hamann, F., Phillips, M. M., Verner, D., Wilkes, B., \& Williams, R. E. 
1996, ApJ, 461, 664

Barlow, T. A. 1993, {\it Ph.D. Dissertation}, University of California 
-- San Diego

Barlow, T. A., \etal 1996, in prep.

Barlow, T. A., Junkkarinen, V. T., Burbidge, E. M., Weymann, R. J., 
Morris, S. L., \& Korista, K. T. 1992, ApJ, 397, 81

Beaver, E. A., \etal 1991, ApJ, 377, L1

Bechtold, J., \etal 1994, AJ, 108, 374

Bergeron, J. 1988, in {\it QSO Absorption Lines: Probing the 
Universe}, eds. J.C. Blades, C. Norman, \& D.A. Turnshek, (Cambridge: 
Cambridge Univ. Press), p. 127

Bergeron, J. \& Stasi\' nska, G. 1986, A\&A, 169, 1

Bica, E. 1988, A\&A, 195, 76

Bica, E., Alloin, D., \& Schmidt 1990, A\&A, 228, 23

Bica, E., Arimoto, N., \& Alloin, D. 1988, A\&A, 202, 8

Blades, J. C. 1988, in {\it QSO Absorption Lines: Probing the 
Universe}, eds. J.C. Blades, C. Norman, \& D.A. Turnshek, (Cambridge: 
Cambridge Univ. Press), p. 147

Chaffee, F. H., Foltz, C. B., Bechtold, J., \& Weymann, R. J. 1986, 
ApJ, 301, 116

Christiani, S., Trentini, S., La Franca, F., Aretxaga, I., Andreani, 
P., Vio, R., \& Gemmo, A. 1996, A\&A, in press

Crenshaw, D. M. 1996, in {\it Emission Lines in Active Galaxies}, 
eds. B.M. Peterson, F.-Z. Cheng and A.S. Wilson, (San Francisco: 
Astr. Soc. Pac.), in press

Fan, X.-M., \& Tytler, D. 1994, ApJS, 94, 17

Ferland, G. J. 1996 ``HAZY, a Brief Introduction to Cloudy'', University 
of Kentucky, Department of Physics and Astronomy, Internal Report

Ferland, G. J., Peterson, B. M., Horne, K., Welsch, W. F., \& 
Nahar, S. N. 1992, ApJ, 387, 95

Ferland, G. J., Baldwin, J. A., Korista, K. T., Hamann, F., Carswell, 
R. F., Phillips, M., Wilkes, B., \& Williams, R. E. 1996, ApJ, 461, 683

Fiore, F., Elvis, M., McDowell, J., Siemiginowska, A., Wilkes, B., 
\& Mathur, S. 1994, ApJ, 431, 515

Foltz, C. B., Chaffee Jr., Weymann, R. J., \& Anderson, S. F. 
1988, in {\it QSO Absorption Lines: Probing the 
Universe}, eds. J. C. Blades, D. A. Turnshek, \& C. A. Norman (Cambridge: 
Cambridge Univ. Press), p. 53

Francis, P. J., Hewitt, P. C., Foltz, C. B., Chaffee, F. H., 
Weymann, R. J., \& Morris, S. L. 1991, ApJ, 373, 465

George, I. M., Turner, T. J., \& Netzer, H. 1995, ApJ, 438, L67

George, I. M., \etal 1996, in prep. 

Gondhalekar, P. M., Kellet, B. J., Pounds, K., Mathews, L., \& 
Quenby, J. J. 1994, MNRAS, 268, 973

Gorgas, J., Efstathiou G., \& Arag\' on Salamanca, A. 
1990, MNRAS, 245, 217

Green ,P. J., \& Mathur, S. 1996, ApJ, 462, 637

Grevesse, N., \& Anders, E. 1989, in {\it Cosmic Abundances of Matter}, 
AIP Conf. Proc. 183, ed. C. I. Waddington (New York:AIP), 1

Halpern, J. P. 1984, ApJ, 281, 90

Hamann, F., Barlow, T. A., Beaver, E. A., Burbidge, E. M., Cohen, 
R. D., Junkkarinen, V., \& Lyons, R. 1995a, ApJ, 443, 606

Hamann, F., Barlow, T. A., \& Junkkarinen, V. 1997a, ApJ, 478, 87 

Hamann, F., Barlow, T. A., R. D., Junkkarinen, V., \& Burbidge, E. M., 
1997b, ApJ, 478, 80

Hamann, F., \& Ferland, G. J. 1992, ApJL, 391, L53

Hamann, F., \& Ferland, G. J. 1993, ApJ, 418, 11

Hamann, F., \& Korista, K. T. 1996, ApJ, 464, 158 

Hamann, F., Korista, K. T., \& Morris, S. L. 1993, ApJ, 415, 541

Hamann, F., Shields, J. C., Ferland, G. J., \& Korista, K. T. 
1995b, ApJ, 454, 688

Hewitt, D. \& Burbidge, G. 1993, ApJS, 87, 451

Jablonka, P., Alloin, D., \& Bica, E. 1992, A\&A, 260, 97

Jannuzi, B. T., \etal 1996, ApJ, 470, L11

Junkkarinen, V. T., Beaver, E. A., Burbidge, E. M., Cohen, R. C., 
Hamann, F., Lyons, R. W., \& Barlow, T. A. 1995, BAAS, 27, 827

Junkkarinen, V. T., Burbidge, E. M., \& Smith, H. E. 1987, ApJ, 317, 
460

Kallman, T. R. \& Mushotzky, R. 1985, ApJ, 292, 49

Kennicutt, R. C., and Garnett, D. R. 1996, ApJ, 456, 504

K\" oppen, J., \& Arimoto, N. 1990, A\&A, 240, 22

Koratkar, A., \etal 1996, ApJ, in press

Korista, K. T., Weymann, R. J., Morris, S. L., Kopko, M., Jr., Turnshek,
D. A., Hartig, G. F., Foltz, C. B., Burbidge, E. M., \& Junkkarinen, V. T.
1992, ApJ, 401, 529

Korista, K. T., Hamann, F., Ferguson, J., \& Ferland, G. J. 1996, 
ApJ, 461, 641

Kriss, G. A., Davidson, A. F., Zheng, W., Kruk, J. W., \& Espey, B. R. 
1995, ApJ, 454, L7

Kriss, G., \etal 1996a, ApJ, in press

Kriss, G., \etal 1996b, ApJ, in press

Krolik, J., \& Kriss, G. 1995, ApJ, 447, 512

Kuhn, O., Bechtold, J., Cutri, R. Elvis, M., \& Rieke, M. 1995, ApJ, 
438, 643

Laor, A., Fiore, F., Elvis, M., Wilkes, B. J., \& McDowell, J. C. 1994, 
ApJ, 435, 611

Lyons, R., \etal 1996, in prep 

Maran, S. P., \etal 1996, ApJ, 465, 733 

Marshall, H. L., Fruscione, A., \& Carone, T. E. 1995, ApJ, 439, 90

Marshall, F. E., \etal 1993, ApJ, 405, 168

Mathews, W. G., \& Ferland, G. J. 1987, ApJ, 323,456

Mathur, S. 1994, ApJ, 431, L75

Mathur, S., Elvis, M., \& Singh, K. P. 1995a, ApJ, 455, L9

Mathur, S., Elvis, M., \& Wilkes, B. 1995b, ApJ, 452, 230

Mathur, S., Wilkes, B., Elvis, M., \& Fiore, F. 1994, ApJ, 434, 493

McWilliam, A., \& Rich, R. M. 1994, ApJS, 91, 749

M\o ller, P., \& Jakobsen, P. 1990, A\&A, 228, 299

M\o ller, P., Jakobsen, P., \& Perryman, M. A. C., 1994, A\&A, 287, 719

Murray, N., Chiang, J., Grossman, S. A., \& Voit, G. M. 1995, ApJ, 451, 498

Nandra, K., \& Pounds, K. A. 1994, MNRAS, 268, 405

Netzer, H. 1990, in {\it Active Galactic Nuclei} (Berlin:Springer-Verlag), 
57

Netzer, H. 1996, ApJ, in press

O'Brian, P., Gondhalekar, P. M., \& Wilson, R. 1988, MNRAS, 233, 801

Osterbrock, D. E. 1989, {\it Astrophysics of Gaseous Nebulae and 
Active Galactic Nuclei}, University Science Press, p. 88

Pettini, M., Smith, L. J., Hunstead, R. W., \& King, D. L. 1994, ApJ, 
426, 79

Petitjean, P., Bergeron, J. \& Puget, J. L. 1992, A\&A, 265, 375

Petitjean, P., Rauch, M., \& Carswell, R. F. 1994, A\&A, 291, 29




Rich, R. M. 1990, ApJ, 362, 604

Sanders, D. B., Phinney, E. S., Neugebauer, G., Soifer, B. T., 
\& Matthews, K. 1989, ApJ, 347, 29

Sargent, W. L., Steidel, C. C., \& Boksenberg, A. 1989, ApJS, 69, 703

Savaglio, S., D'Odorico, S., \& M\o ller, P. 1994, A\&A, 281, 331

Searle, L., \& Sargent, W. L. W. 1972, ApJ, 173, 25

Shields, G. 1996, ApJL, in press

Shields, J. C., \& Hamann, F. 1996, ApJ, submitted

Shull, J. M., \& Sachs, E. R. 1993, 416, 536

Steidel, C. C. 1990, ApJS, 74, 37

Stocke, J., Morris, S., Weymann, R., and Foltz, C. 1992, ApJ, 396, 487

Tinsley, B. 1980, {\it Fund. of Cosmic Phys.}, 5, 287

Tripp, T. M., Lu, L., \& Savage, B. D. 1996, ApJS, in press


Turner, T. J., \& Pounds, K. A. 1989, MNRAS, 240, 833

Turnshek, D. A.\ 1988, {\it QSO Absorption Lines: Probing the 
Universe}, eds. J.C. Blades, C. Norman, \& D.A. Turnshek, (Cambridge: 
Cambridge Univ. Press), p.17

Turnshek, D. A., Kopko, M., Monier, E., Noll, D., Espey, B., 
\& Weymann, R. J. 1996, ApJ, 463, 110


Ulrich, M. H. 1988, MNRAS, 230, 121

Vila-Costas, M. B., \& Edmunds, M. G. 1993, MNRAS, 265, 199

Voit, G. M., Shull, J. M., \& Begelman, M. C. 1987, 316, 573

Voit, G. M., Weymann, R. J., \& Korista, K. T.\ 1993, ApJ 413, 95

Walter, R., Orr, A., Courvoiser, T. J.-L., Fink, H. H., Makino, F., 
Otani, C., \& Wamsteker, W. 1994, A\&A, 285, 119

Wampler, E., J., Bergeron, J., \& Petitjean, P. 1993, A\&A, 273, 15


Weymann, R. J., Turnshek, D. A., \& Christiansen, W. A. 1985, 
in {\it Astrophysics of Active Galaxies and Quasi-Stellar Objects}, 
ed. J. Miller, (Mill Valley, CA: University Science Books), 185

Weymann, R. J., Morris, S. L., Foltz, C. B., \& Hewett, P. C. 1991,
ApJ, 373, 23 (WMFH)

Weymann, R. J. Williams, R. E., Peterson, B. M., \& Turnshek, 
D. A. 1979, ApJ, 218, 619

Wheeler, J. C., Sneden, C., and Truran, J. W., 1989, ARA\&A,
27, 279 

Wilkes, B. J., \& Elvis, M. 1987, ApJ, 323, 243

Wilkes, B. J., Tananbaum, H., Worrall, D. M., Avni, Y., Oey, M. S., 
\& Flanagan, J. 1994, ApJS, 92, 53

Worthey, G., Faber, S. M., \& Jes\' us Gonzalez, J. 1992, ApJ, 398, 69

York, D. G., Yanny, B., Crotts, A., Carilli, C., Garrison, E., 
\& Matheson, L. 1991, MNRAS, 250, 24

Zamorani, G., \etal 1981, ApJ, 245, 357

\bigskip\bigskip
\begin{center}
{\bf Figure Captions}
\end{center}
\leftskip=1.5em
\parindent=-1.5em

\underline{Figure 1.} Synthetic QSO spectra used in the calculations 
are shown normalized to unity at 2500~\AA . Panels 
A, B and C show the continuum types A, B and C (see text \S2.1). 
The solid curves illustrate a range of plausible shapes 
allowed by the free parameters \aox\ plus log~$T_c$ (for continuum A) 
or \aox\ plus log~$E_c$ (for B and C). 
The bold solid lines indicate the fiducial cases A(5.7,$-$1.6), 
B(1.4,$-$1.6) and C(1.4,$-$1.6). The dotted line labeled MF87 in 
each panel is the generic AGN spectrum from Mathews \& Ferland (1987). 
Panel D shows the MF87 continuum and the fiducial cases 
for A, B and C after transmission through an X-ray warm absorber 
(\S2.1). 
The transmitted spectra of types A, B and C are labeled 
AT, BT, and CT, respectively, in panel D.  
\bigskip

\underline{Figure 2a.} Ionization fractions, $f$(M$_i$) and $f$(HI), are shown 
for different ionization parameters, $U$, in clouds that are 
optically thin at all continuum wavelengths. 
The ionizing spectra are A(5.7,$-$1.6) -- left window -- and B(1.4,$-$1.6) 
-- right window. The $f$(HI) values are shown across the top.  
The various $f$(M$_i$) curves are labeled at their peaks. The two dotted 
curves in the top panels are S4 (not labeled) and S6. The two dash-dot 
curves in those panels are Fe2 (not labeled) and Fe3. The two 
dash-dot curves in the middle panels are Al2 (not labeled) and 
Al3. The three dash-dot curves in the bottom panels are 
Si2, Si3 (not labeled) and Si4. 
\bigskip

\underline{Figure 2b.} C(1.4,$-$1.6) and MF87. See Fig. 2a.
\bigskip

\underline{Figure 2c.} PL($-$1.0) and PL($-$1.5). See Fig. 2a.
\bigskip

\underline{Figure 3a.} A(5.2,$-$1.6) and A(6.0,$-$1.6). See Fig. 2a.
\bigskip

\underline{Figure 3b.} A(5.7,$-$1.3) and A(5.7,$-$1.9). See Fig. 2a.
\bigskip

%

\underline{Figure 4a.} C(1.1,$-$1.6) and C(1.9,$-$1.6). See Fig. 2a.
\bigskip

\underline{Figure 4b.} C(5.7,$-$1.3) and C(5.7,$-$1.9). See Fig. 2a.
\bigskip

%

\underline{Figure 5.} Continuum A; contours of constant ionization correction 
normalized to solar,  
log($f$(HI)/$f$(M$_i$))~+~log(H/M)$_{\odot}$, evaluated at the 
peaks in the $f$(M$_i$) curves are plotted for optically thin 
clouds photoionized by various continua with the parameters $T_c$ and 
\aox\ as shown. Each window applies to a different ion, labeled at 
the top. The dotted contours appear at 
increments of 0.1 dex, the thin solid contours appear every 0.25 dex 
and the bold contours every 1.0 dex. See \S3.1. 
\bigskip

\underline{Figure 6.} Continuum C; normalized ionization corrections analogous  
to Fig. 5. 
\bigskip

\underline{Figure 7.} Ionization corrections normalized by solar abundances, 
log($f$(HI)/$f$(M$_i$))~+~log(H/M)$_{\odot}$, for optically thin 
clouds are shown for different ionization parameters, $U$. 
The ionizing spectra are A(5.7,$-$1.6) at left and C(1.4,$-$1.6) right. 
The curves are labeled at their minima as in Fig. 2a.
\bigskip

\underline{Figure 8.} Continuum A; minimum normalized ionization corrections,
log($f$(HI)/$f$(M$_i$))~+~log(H/M)$_{\odot}$, evaluated at the 
minima in the $f$(HI)/$f$(M$_i$) curves, are plotted for various ions 
in optically thin clouds. The format follows Fig. 5. 
\bigskip

\underline{Figure 9.} Continuum C; minimum normalized ionization corrections  
analogous to Fig. 8.
\bigskip

\underline{Figure 10a.} Continuum A; normalized ionization corrections, 
log($f(b_j)$/$f(a_i)$)~+~log($b$/$a$)$_{\odot}$, for deriving 
relative metal abundances [$a$/$b$] from Eqn. 2. The 
curve labeled N5/C4 gives the correction 
factor needed to derive [N/C] from the ratio of N V/C IV column 
densities, etc. The curve for P5/C4 is shifted downward by $-$3.0 
for convenience. Each panel 
shows results for a different ionizing continuum shape. 
\bigskip

\underline{Figure 10b.} Continuum C; normalized ionization corrections 
for deriving relative metal abundances, analogous to Fig. 10a.
\bigskip

\underline{Figure 11.} Ionization fractions for optically thin clouds ionized 
by continuum CT (transmitted through a warm absorber; left-hand 
panel), and for clouds with 
$\log~N$(HI)~$\leq$~16.5~\cmsq , $\log~N_H$~$\leq$~22.0~\cmsq , 
solar abundances and ionized by the fiducial continuum C (right panel). 
The format follows Fig. 2a. 

\end{document}